\documentclass[twocolumn,floatfix,showpacs,preprintnumbers,amsmath,amssymb]{revtex4}
\usepackage{graphicx}
\usepackage{subfigure}
\usepackage{amssymb}
\usepackage{amsmath}
\usepackage{bm}
\begin{document}

\title{Surface Gap Solitons in Exciton Polariton Condensates}

\author{Ting-Wei Chen$^{1}$}
\author{Szu-Cheng Cheng$^{2}$}
\email {sccheng@faculty.pccu.edu.tw}
\thanks{FAX: +886-2-28610577}

\affiliation{$^{1}$Department of Electrophysics, National Chiayi University, Chiayi city 60004, Taiwan, R. O. C.\\
$^{2}$Department of Optoelectric Physics, Chinese Culture University, Taipei 11114, Taiwan, R. O. C.}

\date{\today}

\begin{abstract}
A gap soliton is a solitonic state existing inside the band gap of an infinite-periodic exciton-polariton condensate (EPC). The combination of surface states and gap solitons forms the so named surface gap solitons (SGSs). We analyze the existence of SGSs near the interface between uniform and semi-infinite periodic EPCs. We find that SGSs exist only when the system is excited by a pump with low power and small width. As the pump power or width increases, SGSs become unstable.
\end{abstract}


\maketitle
Solitons - nonspreading localized waves \cite{soliton} are created when the nonlinear focusing dynamics compensates the spreading due to dispersions.  The bright gap soliton, which is a solitonic state existing inside the band gap of infinite-periodic systems, was observed in periodic photonic \cite{18,opticalsoliton1} and matter-wave \cite{brightsoliton2,soliton1,brightsoliton1} band structures.

In recent years, exciton-polariton condensates (EPCs) created in semiconductor microcavities \cite{2} have been the subject of intensive research because of their potential advances towards a new generation of low-threshold lasers and ultrafast optical amplifiers and switches at room temperature \cite{3}.  The EPC in a static \cite{13,14} or tunable \cite{15,16} periodic potential has been realized to study the s- and p-type wavefunctions which could have different energies, symmetry and spatial coherence. Understanding the properties of EPCs in a periodic potential is of importance towards developing quantum simulation on which most researchers have been working with cold atoms or ion traps \cite{quantumsimulation}.  The solitonic gap states were observed by using a laser pump aimed at the barrier region of the one-dimensional (1D) periodic potential \cite{17}, which was produced by laterally modulated micro-metal wires.  The resonantly excited gap solitons were also observed in a two-dimensional square lattice \cite{Gorbach,2Dsoliton}.  A mechanism for self-localization of EPCs was theoretically proposed due to effective potentials induced by the exciton-polariton flows \cite{self-localization}.  Dark gap solitons were also shown to be possibly occurred in EPCs \cite{cheng2018}.  Nevertheless, the studies of surface solitonic modes, named surface gap solitons (SGSs), in EPCs are scarce nowadays.   SGSs are polaritonic waves that combine surface states and gap solitons into nonlinear localized states near the interface between uniform and semi-infinite periodic EPCs.  The polaritonic SGSs can be in analogy to the optical nonlinear Tamm states \cite{Tamm1,Tamm2,Tamm3,Tamm4}.

In this paper, the effects of spatial localized SGSs of EPCs are investigated by exciting the system via pumps with various powers and widths.  We first adjust the uniform and semi-periodic potential strengths.  Then we tune the excitation energy into the energy-band gap formed in a periodic EPC to find the steady-state SGSs.  We demonstrate that there are stable SGSs near the band edge with anomalous dispersion relation in the first Brillouin zone.  The anomalous dispersion relation could produce a focusing nonlinearity in the formation of a bright gap soliton in EPCs \cite{Egorov2009}.  It is noteworthy that the SGS becomes unstable and detaches itself from the interface while the pump strength or width is increasing.

In the theoretical modeling of the EPC, we rely on the mean-field complex Gross-Pitaveskii equation (cGPE) incorporated the external potential, inter-particle interactions, pump and loss \cite{Keeling}:
\begin{equation}\label{1}
i\hbar\frac{\partial \Psi}{\partial t}=-\frac{\hbar^{2}}{2m}\frac{\partial^{2}\Psi}{\partial{x}^{2}}+\tilde V(x)\Psi+U|\Psi|^{2}\Psi+i(\gamma_{eff}-\Gamma|\Psi|^{2})\Psi,
\end{equation}
where $\Psi$ is the wave function, $\hbar$ and $m$ are Planck's constant and exciton-polariton mass, respectively.  $\tilde V(x)$ is a one-dimensional potential given by $\tilde V(x)=\tilde V_{1}sin^{2}(\pi x/a)$ and $\tilde V(x)=\tilde V_{2}$ if $x\ge0$ and $x<0$, respectively.  Here $\tilde V_{1}$ and $\tilde V_{2}$ are the external potential strengths; $a$ is the lattice constant of periodicity.  The third term on the right-hand side of the equation represents the normalized two-body interaction with $U$ being the strength of the two-body interaction potential.  $\gamma_{eff}$ represents the linear net gain describing the balance between the stimulated scattering of exciton-polaritons into the condensate and the linear loss of exciton-polaritons out of the cavity.  $\Gamma$ is the coefficient of gain saturation.  Interested readers can refer to the article by Szyma$\acute{n}$ska, Keeling and Littlewood \cite{GPE} for how Eq.(\ref{1}) is obtained.

Let the time scale be $1/\omega$, where $\omega=4\hbar/ma^{2}$.  Choosing lengths in units of $a/2$, energies in units of $\hbar\omega$ and the wave function $\Psi\rightarrow\sqrt{\hbar\omega/U}\Psi$ with respect to $U$, the steady states of Eq.(\ref{1}) are obtained by taking the wavefunction $\Psi(x,t)=\psi(x)e^{-iEt}$, where $E$ is the chemical potential or excitation energy of the system.  Then Eq.(\ref{1}) becomes
\begin{equation}\label{2}
\frac{1}{2}\frac{\partial^{2}\psi}{\partial{x}^{2}}+(E-V(x)) \psi-|\psi|^{2}\psi-i(\alpha(x)-\sigma|\psi|^{2})\psi=0,
\end{equation}
where $\sigma=\Gamma/U$.  In this paper $\sigma=0.52$ is chosen so that the energy difference between point "p" and point "z" in Figure 1 is 1 meV that is shown in Ref.\cite{13}.  The external potential $V(x)$ is given by $V(x)=V_{1}sin^{2}(\pi x/2)$ and $V(x)=V_{2}$ in the regions of $x\ge0$ and $x<0$, respectively, where $V_{1}=\tilde V_{1}/\hbar\omega$ and $V_{2}=\tilde V_{2}/\hbar\omega$.  We consider the pump with a finite width, i.e., $\alpha(x)=\alpha_{0} e^{-(x/w)^2}$ with pump power $\alpha_{0}=\gamma_{eff}/\hbar\omega$ and width $w$.

Before solving Eq.(\ref{2}) to find the steady states of SGSs, we would like to obtain the linear band structure of an infinite periodic potential and its corresponding Bloch states.  Due to the external potential being composed by uniform and semi-periodic potentials, the wavefunction $\psi(x)$ is then divided into two parts: $\psi(x)=\psi_{<}(x)$ and $\psi(x)=\psi_{>}(x)$ in the regions of $x<0$ and $x\ge0$, respectively.  Ignoring the nonlinear, loss and pump effects, the Bloch wavefunction $\chi(x)$ is given by the form $\chi_{n,k}(x)=e^{ikx}u_{_{n,k}}(x)$, where $k$ is the quasi-momentum and $n$ indicates the band index.  The Bloch functions $u_{n,k}(x)$ are periodic with period $2$, i.e., $u_{n,k}(x+2)=u_{n,k}(x)$.  In this case, the function $u_{n,k}(x)$ is expanded by a Fourier series in a reciprocal momentum space, i.e., $u_{n,k}(x)=\displaystyle\sum\limits_{m}C^{n}_{m}e^{imGx}$, where $C^{n}_{m}$ are the expansion coefficients of the Bloch function and $G=\pi$ is the reciprocal wave vector.  Therefore, the linear spectra of an EPC consists of bands of eigenenergies $E_{n}(k)$.

\begin{figure}
\centering
\scalebox{.24}{\includegraphics{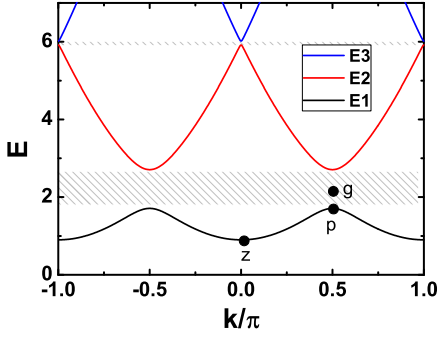}}
\caption{(Color online) Linear band structure without pump and loss effects for $V_{1}=2$.  Surface gap solitons occur near the band edge and inside the second band gap labeled by "p" and "g" points, respectively.}
\end{figure}

\begin{figure}
\centering
\subfigure{\scalebox{.15}{\includegraphics{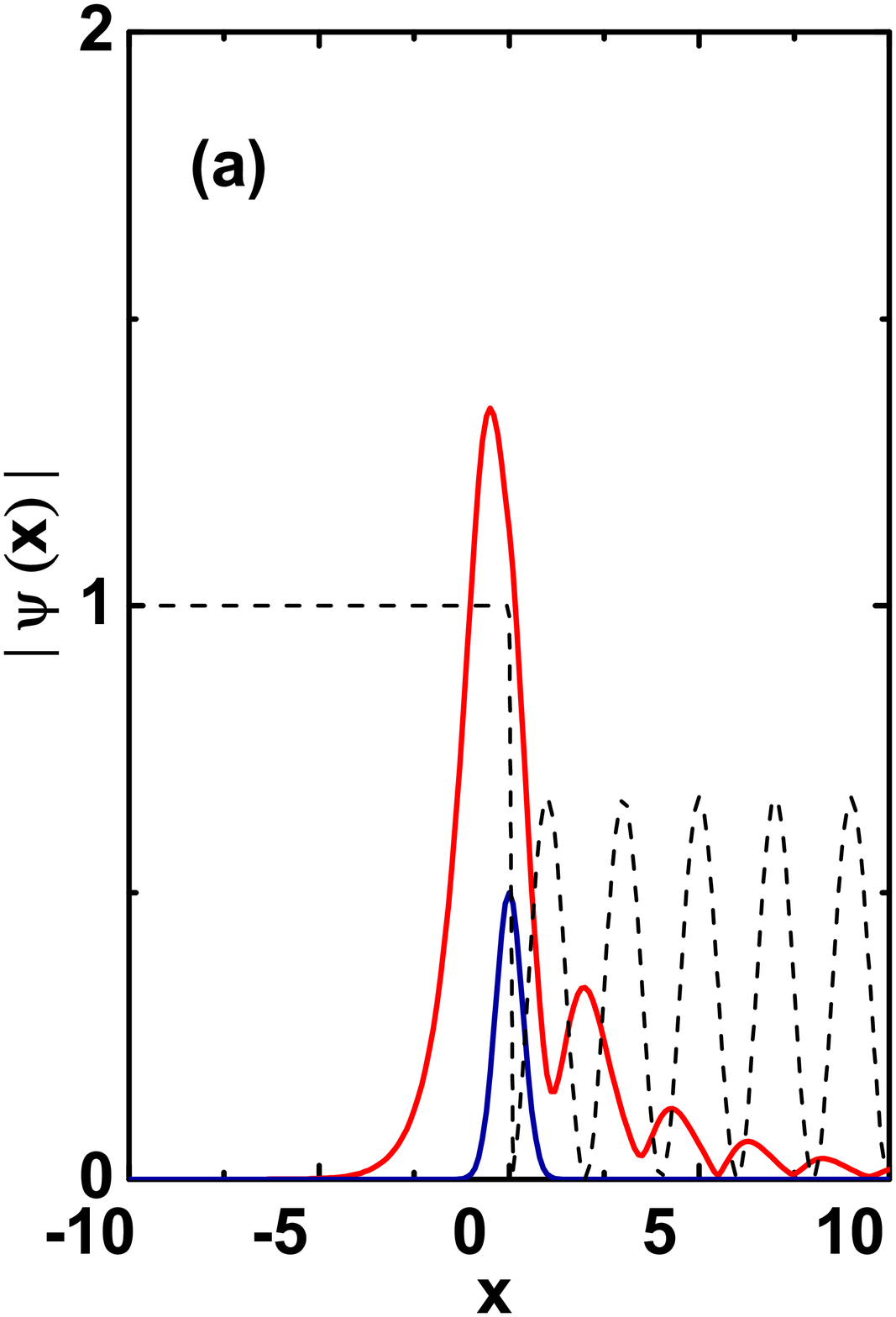}}}
\subfigure{\scalebox{.15}{\includegraphics{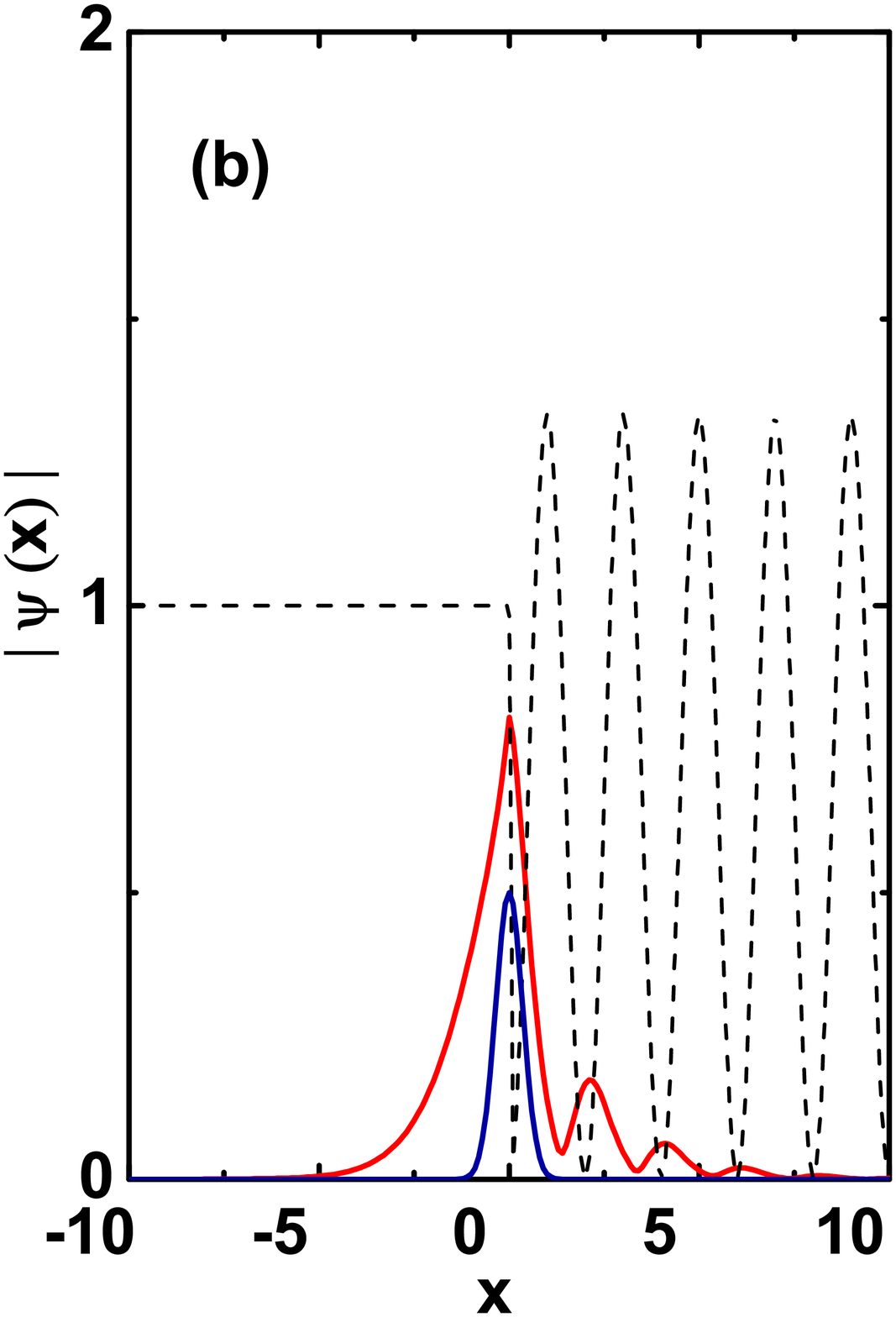}}}\\
\caption{Surface gap solitons (red solid curves) originating from the second band gap for two pump strengths (a) $V_{2}=2$ and (b) $V_{2}=4$. Here, $\delta=0.4$, $V_{1}=2$, $\alpha_{0}=0.5$, and $w=0.5$.  The black dash and blue solid curves are the relative strength of the external potential and pump profiles, respectively.}
\end{figure}

Figure 1 shows the energy band diagram with $V_{1}=2$ neglecting the nonlinear, pump and loss effects.  The existence of the periodic potential splits the energy dispersion into multiple bands with the band edge occurred at $k=\pi/2$.  Gaps between the bands are getting wider for a deeper potential depth (larger $V_{1}$).  The lowest and intermediate energy bands are the first and second bands, respectively.  There are two open gaps below the first band center at $k=0$ and above the second band edge at $k=\pi/2$.  The energy curvature is negative near the band edge labeled by point "p" shown in Fig. 1.  The band edge energies of the first and second bands are $E_{1}(k=\pi/2)=1.709$ and $E_{2}(k=\pi/2)=2.709$, respectively.  We would like to analyze the existence of the SGS state originating from the energy gap labeled by point "g" (see Fig. 1).

After finding the Bloch bands and its corresponding wavfunctions, we apply the effective-mass approximation\cite{cheng2018,EMT,MSM} to Eq.(\ref{2}).  The nonlinear and non-equilibrium effects are treated as the perturbations.  A wave packet $\psi_{>}(x)$ with a small momentum distribution centered around $k_{0}=\pi/2$ in the $n=1$ band is described by a slowly varying envelope $F(x)$ (on the scale of several lattice constants) multiplied by the Bloch wavefunction as $\psi_{>}(x)=F(x)\chi_{1,k_{0}}(x)$.  In the case of weakly interacting polaritons and negligible band mixing, the steady-state envelop $F(x)$, with energy $E$ near the band edge at $k_{0}$, satisfies the time-independent differential equation written as
\begin{multline}\label{3}
\frac{-1}{2|m^{*}(x)|}\frac{d^{2}F(x)}{d{x}^{2}}+\delta F(x)-\lambda |F(x)|^{2}F(x)\\-i(\alpha(x)-\sigma\lambda |F(x)|^{2})F(x)=0,
\end{multline}
where $1/m^{*}=\frac{\partial^{2}E_{1}(k)}{\partial{k}^{2}}|_{k=k_{0}}$ is the inverse effective mass of the exciton-polariton. $\delta=E-E_{1}(k_{0})$ is the detuning energy above the band edge of the first band.  Due to the periodicity, the interaction strength between exciton-polaritons is changed and renormalized by a factor $\lambda$, where $\lambda=\int_{-1}^{1}|\chi_{1,k_{0}}(x)|^{4}dx/\int_{-1}^{1}|\chi_{1,k_{0}}(x)|^{2}dx$.  The periodic potential determines the energy bands and effective masses through the formalism of Bloch wavefunctions.  It also leads to a group velocity $v_{g}$ of the wave packet determined by the energy band via $v_{g}(k)=\frac{\partial E_{1}(k)}{\partial{k}}$. We have $v_{g}(k_{0})=0$ and $m^{*}<0$ here.  In EPCs with a negative effective mass, a bright soliton can be achieved due to the repulsive nonlinearity in EPCs being reversed into an attractive nonlinearity that can then balance the dispersion effect \cite{Gorbach,2Dsoliton}.

Eq.(\ref{3}) determines the envelope of the steady wavefunction $\psi_{>}(x)$ in the region of $x\ge0$.  The steady wavefunction $\psi_{<}(x)$ in the region of $x<0$ is determined by Eq.(\ref{3}) with $V(x)=V_{2}$.  Still, there are no simple analytic solutions of Eqs.(\ref{2}) and (\ref{3}).  We have to rely on the Newton-Ralphson method (NRM) to solve both equations numerically. Applying the NRM, we need an initial wavefunction to generate a self-consistent solution of a SGS.  A toy-model equation \cite{cheng2018} and its analytic solution, $\phi_(x)$, could be obtained if $|\psi_{<}(x)|^{2}$ in Eq.(\ref{2}) and $|F(x)|^{2}$ in Eq.(\ref{3}) are replaced by $\psi_{<}(x)^{2}$ and $F(x)^{2}$, respectively.  We then use $\phi_(x)$ as the initial wavefunction of the NRM to solve Eqs.(\ref{2}) and (\ref{3}) consistently.  The wavefunction $\phi_(x)$ of the toy model is determined by the following equation:
\begin{multline}\label{4}
\frac{-1}{2|M(x)|}\frac{d^{2}\phi(x)}{d{x}^{2}}+\Delta(x) \phi(x)-\Lambda(x) \phi(x)^{3}\\-i(\alpha(x)-\sigma\Lambda(x)\phi(x)^{2})\phi(x)=0,
\end{multline}
where $M(x)=m^{*}$, $\phi_{>}(x)=F(x)$, $\Delta(x)=\delta$, and $\Lambda(x)=\lambda$ in the region of $x\ge0$. In the region of $x<0$, $M(x)=1$, $\phi_{<}(x)=\psi_{<}(x)$, $\Delta(x)=\epsilon$, and $\Lambda(x)=1$, where $\epsilon=E-V_{2}$.

We could find the analytic solution of Eq.(\ref{4}) if the pump profile $\alpha(x)$ is treated as a uniform pump with power $\alpha_{0}$, i.e., $\alpha(x)=\alpha_{0}$.  In this paper, we consider the situation that the excitation energy $E$ is higher or less than the band-edge energy $E_{1}(k_{0}=\pi/2)$ or potential energy $V_{2}$, respectively.  Then Eq.(\ref{4}) with $\delta>0$ and $\epsilon<0$ contains two analytic solutions given by a hyperbolic secant solution $\phi_{>}(x)=C_{1}$sech($B_{1}x+\theta_{1}$) and hyperbolic cosecant one $\phi_{<}(x)(x)=C_{2}$csch($B_{2}x+\theta_{2}$) in the regions of $x\ge0$ and $x<0$, respectively.  Here, $B_{1}=\sqrt{2|m^{*}|(\delta-i\alpha_{0})}$, $C_{1}=\sqrt{2(\delta-i\alpha_{0})/(\lambda(1-i\sigma))}$, $B_{2}=\sqrt{2(|\epsilon|+i\alpha_{0})}$ and $C_{2}=\sqrt{2(|\epsilon|+i\alpha_{0})/(1-i\sigma)}$.  The phases $\theta_{1}$ and $\theta_{2}$ are determined by the boundary conditions that $\Phi_{>}(x)|_{x=0}=\phi_{<}(x)|_{x=0}$ and $\frac{d \Phi_{>}(x)}{d{x}}|_{x=0}=\frac{d \phi_{<}(x)}{d{x}}|_{x=0}$, where $\Phi_{>}(x)=\phi_{>}(x)\chi_{1,k_{0}}(x)$.  The analytical hyperbolic solutions provides qualitative understanding that the SGS is self-localized near the interface between uniform and semi-infinite periodic potentials.

After finding the analytic solution of Eq.(\ref{4}), wavefunctions $\Phi_{>}(x)$ and $\phi_{<}(x)$ are treated as the initial wavefunction $\psi(x)$ in the NRM of solving Eq.(\ref{2}) to obtain a SGS numerically.  The self-consistent solutions of SGSs are shown in Fig. 2.  The SGS density distributions for two different $V_{2}$ values are shown by red solid curves.  The density distributions monotonically diminish from the interface between uniform and semi-infinite periodic potentials towards lef-hand and right-hand sides.  Due to the energy detuning $\epsilon<0$, the uniform potential on the left-hand side of the interface creates a barrier to prevent the SGS from delocalizing.  Still, there are some exciton-polaritons penetrating the barrier and accumulating near the left-hand side of the interface.  On increasing the pump strength, the maxima density peak from accumulated exciton-polaritons is growing higher and shifts towards the interface.  On the right-hand side of the interface, the EPC density displays an oscillating and gradually diminishing distribution.  The crests of the oscillating density are centered on the periodic potential minima due to the expulsive force from the potential.  The band gap from the periodic potential forbids the EPC to propagate toward the right-hand side if $\delta>0$.  Finally, with the disperse effect being balanced by the attractive nonlinearity, a SGS is achieved due to the existence of a negative effective mass ($m^{*}<0$) for the right-hand side EPC.  The density profiles of SGSs would change if the periodic potential depth is higher than the uniform potential depth, i.e., $V_{1}>3$ shown in Fig. 2(b).  The maxima-density peak of a SGS tilts towards the interface and the number of crests of the oscillating density is reduced.

\begin{figure}
\centering
\scalebox{.24}{\includegraphics{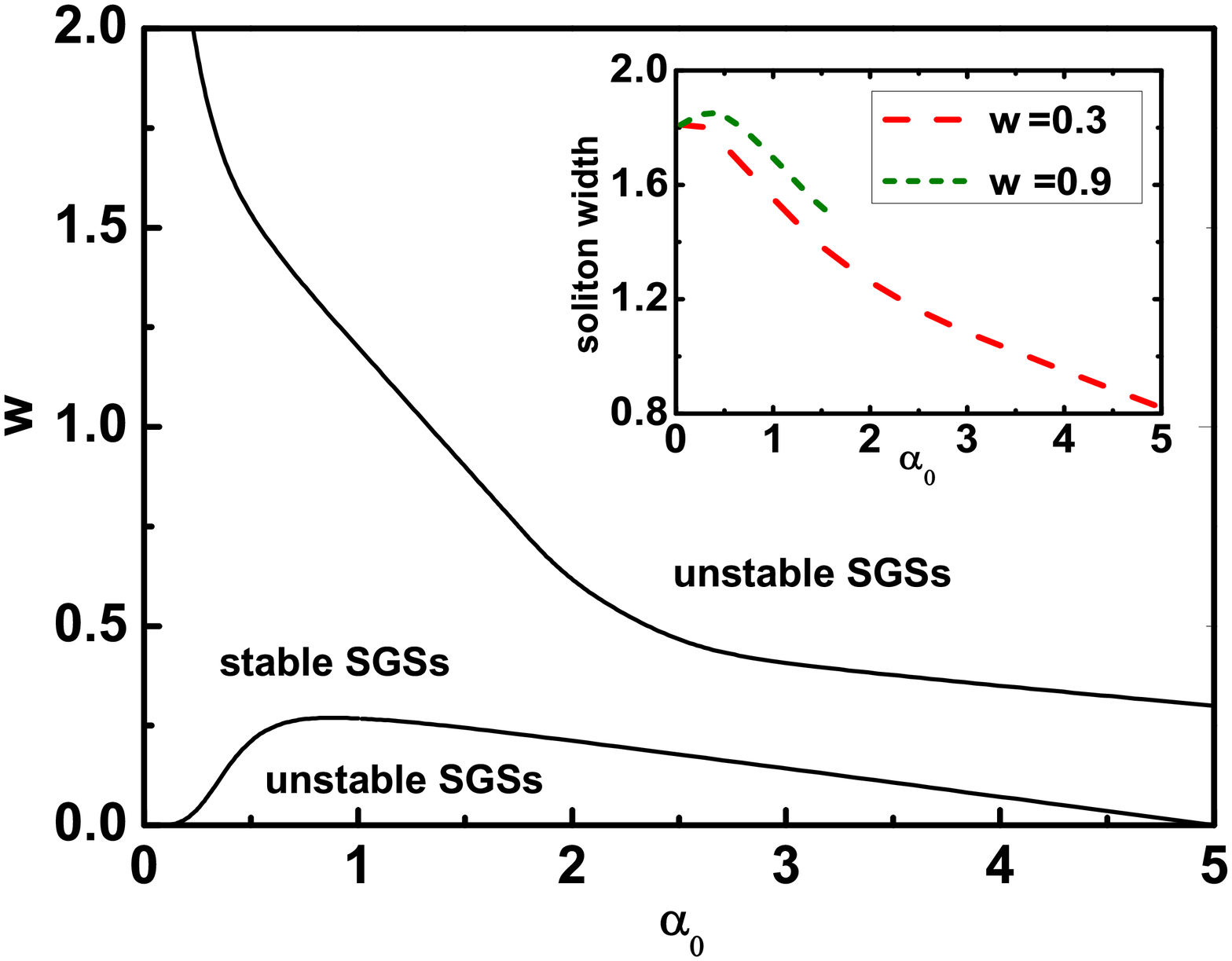}}
\caption{(Color online) Phase boundaries between stable and unstable surface gap solitons.  Boundaries are plotted for various pump powers and widths.  The physical parameter of $\delta$, $V_{1}$ and $V_{2}$ are 0.4, 2 and 3, respectively.  Choosing various pump widths, insets are soliton widths versus pump strengths.}
\end{figure}

\begin{figure}
\centering
\scalebox{.24}{\includegraphics{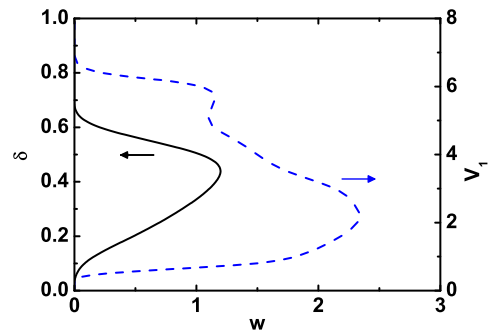}}
\caption{(Color online) Phase boundaries between stable and unstable surface gap solitons.  Stable phases are on the left-hand of curves.  The black solid line is the boundary for various energy detunings and pump widths under $\alpha_{0}=1$, $V_{1}=2$ and $V_{2}=3$.  The blue dash line is the boundary for various periodic potential depths and pump widths under $\alpha_{0}=1$, $\delta=0.4$ and $V_{2}=3$.}
\end{figure}

The magnitudes of $V_{1}$ and $\delta$ have minor roles to the formation of a SGS.  Instead, the pump profile is the dominant factor to determine the instability of a SGS.  Bogoliubov excitations can be studied by inserting into rescaled Eq.(\ref{1}) a solution that has small perturbations to the steady wavefunction \cite{excitation}.  Then, $\Psi(x,t)=e^{-iEt}(\psi(x)+u_{q}(x)e^{iqx}e^{-i\Omega t}-v^{*}_{q}(x)e^{-iqx}e^{i\Omega t})$, where $u_{q}(x)$ and $v_{q}(x)$ are the amplitudes of Bogoliubov quasi-particles with wavevector $q$.  Substituting $\Psi(x,t)$ into rescaled Eq.(\ref{1}), we can obtain two coupled Bogoliubov-equations to judge the stability of the SGSs through the complex-valued $\Omega$.  For various pump powers and widths, the phase boundary between stable and unstable SGSs in the first band gap (between the first and second bands) are shown in Fig. 3.  SGSs become unstable in regimes with higher pump powers and larger pump widths or with very narrow pumps.  For a large pump width, a SGS owns a much broad density distribution which can be easily delocalized away from the interface.  Therefore, a SGS under a uniform pump is unstable.  A SGS under a very narrow pump does not have enough density to balance the dispersion effect and become stable.  In a given pump, stable SGSs exist when the pump owns an intermediate width.  On the right-hand side of the interface, the number of density sidelobes of a stable SGS is unaffected by the pump power.  On the other hand, the width of the highest density peak decreases as the pump power increases (see insets in Fig. 3).

While fixing $\alpha_{0}=1$ and $V_{2}=3$, two new phase boundaries are drawn in Fig. 4.  One is the boundary, shown by the blue dash line, for various periodic potential depths and pump widths under $\alpha_{0}=1$, $\delta=0.4$ and $V_{2}=3$.  Another is the boundary, shown by the black solid line, for various energy detunings and pump widths under $\alpha_{0}=1$, $V_{1}=2$ and $V_{2}=3$.  Stable phases are on the left-hand of curves.  From the blue dash line, we conclude that a shallow semi-periodic potential does not create a large enough band-gap to localize a SGS.  There is also no localized SGS in a deeper semi-periodic potential which produces a band-edge energy higher than the uniform potential barrier, i.e., $\epsilon>0$.  Near the first band edge with low energy detuning inside the gap, a SGS owns a much broad density distribution and is easily delocalized away from the interface (see the black solid line in Fig. 4).  On the other hand, moving the detuning close to the second band edge, the nonlinearity of an EPC gradually changes from the attractive nonlinearity into the repulsive nonlinearity.  No stable SGS with repulsive nonlinearity can exist in EPCs.

In summary, we showed that stable SGSs could be created near the interface between uniform and semi-infinite periodic EPCs.  Periodicity induces a nonlinear character change.  Near the first band edge, an EPC owns a negative effective mass and its repulsive nonlinearity is reversed into an attractive nonlinearity that could balance the dispersion effect.  We also reported the dynamical instability of SGSs. We found that SGSs become unstable as the pump power and width are increasing.  No stable SGSs exist in a uniformly pumped EPC.

We acknowledge the financial support from Ministry of Science and Technology of the Republic of China under contract No. MOST105-2112-M-034-001-MY3 and MOST105-2112-M-415-009-MY3.

\bibliography{SGS}

\end{document}